# Engineering the Collapse of Lifetime Distribution of Nitrogen-Vacancy Centers in Nanodiamonds

H. Li[1,*], J. Y. Ou[1], B. Gholipour[1], J. K. So[2], D. Piccinotti[1], V. A. Fedotov[1], N. Papasimakis[1]

[1] *Optoelectronics Research Center & Center for Photonic Metamaterials, University of Southampton, Southampton SO17 1BJ, UK*

[2] *Center for Disruptive Photonic Technologies, School of Physical and Mathematical Sciences and The Photonics Institute, Nanyang Technological University, Singapore 637371, Singapore*

[*] H.Li@soton.ac.uk

**ABSTRACT**

We demonstrate experimentally that the distribution of the decay rates of nitrogen-vacancy centers in diamond becomes narrower by over five times for nanodiamonds embedded in thin chalcogenide films.

**Keywords:** diamond NV centers, cathodoluminescence, spontaneous emission rate, lifetime, chalcogenide

The nitrogen-vacancy (NV) center is a point defect in diamond formed by a lattice vacancy paired with a nitrogen atom, which replaces the carbon atom adjacent to the vacancy.[1] NV centers are fluorescent in the red part of the optical spectrum and, individually, can act as well-localized sources of single photons with high photostability even at room temperature.[2-5] The intensity and spectrum of NV emission are highly sensitive to external perturbations, such as temperature, strain, electric and magnetic fields.[1,6] Although naturally present in large single-crystal diamonds, NV centers have been routinely produced in nanocrystals (nanodiamonds), which allow efficient extraction of NV fluorescence and enable integration with optical microresonators, thus holding promise as an accessible and cost efficient optical quantum platform for sensing, communication and computing.[7-12] NV centers in nanodiamonds (NDs) can also be used as scanning-probes for fluorescence lifetime imaging microscopy,[13] especially given their good bio-compatibility.[14] Yet, real-life applications of NV centers in NDs has been difficult in view of the broad distribution of their lifetimes. The broadening results from strong variations of both radiative and nonradiative decay rates, and quantum efficiency of NV centers,[15] which in practice are controlled by many factors, such as crystal strain,[15] contamination during fabrication,[16] coupling to Mie resonance of NDs,[17-18] locations and orientations of NV centers within NDs,[19-20] and the presence of large ensembles of NV centers in NDs.[21]

In this Letter, we report that the statistics of the decay rates of NV centers can be efficiently controlled by changing the immediate dielectric environment of NDs. We demonstrate this experimentally by characterising the lifetime distribution for large ensembles of NV centers in NDs embedded in thin chalcogenide films, the optical constants of which can be tuned continuously from plasmonic to lossy dielectric. Our analysis indicates that the distribution of lifetimes can become narrower by over five times, featuring a spread of 1 ns around the average lifetime of 11 ns.

The NDs in our study had an average size of 120 nm and contained ~1200 $NV^0$ (neutral) centers each (obtained from Sigma Aldrich). As a host for NDs we employed a binary alloy of antimony telluride (SbTe), which is an archetypal chalcogenide. NDs were embedded in SbTe using the following procedure. First, a thin film of SbTe of varying composition and ranging in thickness, $d/2$, from 20 to 50 nm was deposited over a 28 mm × 28 mm large area of silicon (Si) substrate using a high-throughput physical vapour deposition system. It was equipped with off-axis Knudsen cell sources symmetrically arranged around the target substrate, which was held under $\leq 10^{-8}$ mbar vacuum at room temperature. The density gradient of Sb and Te (each of ≥ 99.9999 % purity) along the substrate was independently controlled using fixed wedge shutters, which ensured that for every SbTe composition (i.e., at every point on the substrate) the atomic components mix simultaneously.[22] During the synthesis of the film, the phase of the deposited SbTe alloy was formed directly as amorphous.[23] The obtained SbTe film was then coated via drop casting with a dispersion of NDs, which was prepared by diluting NDs in methanol and mixing the solution for 10 min in an ultrasonic bath. After methanol evaporated, a second SbTe film (with thickness $d/2$) was deposited over the first SbTe film using the same procedure as above, which yielded an SbTe film with overall thickness $d$ incorporating NDs. Figure 1(a) shows a fragment of the sample featuring ND clusters covered by SbTe. As a reference, we also prepared a sample containing exposed NDs, where the dispersion of NDs was applied onto an SbTe film of thickness $d$, i.e. after the second round of the deposition. All samples were vacuum sealed immediately upon production, and also between characterisation measurements.

The complex permittivity of the synthesized SbTe films was a strong function of composition. It was measured using a variable angle spectroscopic ellipsometer (J. A. Woollam M2000) with an automated stage programmed to sample a $10 \times 10$ point array within the central 19 mm × 19 mm region (point-to-point separation 2.1 mm). Figures 1(b) and 1(c) show the real, Re $\varepsilon$, and imaginary, Im $\varepsilon$, parts of SbTe permittivity mapped across the samples at a wavelength of 575 nm (which corresponds to the zero phonon line of $NV^0$ centers). Re $\varepsilon$ is seen to vary from –23 to 13 crossing zero for a wide range of compositions. Im $\varepsilon$ remains fairly high for all compositions, in the range 15 – 30. Variation in the thickness of the obtained SbTe films is also inherent to the process of synthesizing films from two off-axis sources with wedge shutters.[23-24] The thickness was measured using a stylus profilometer along the external edges of the film and then interpolated across the central area. Figure 1(d) reveals that the thickness of SbTe films, $d$, obtained after two rounds of deposition varied across the samples from about 40 to 100 nm.

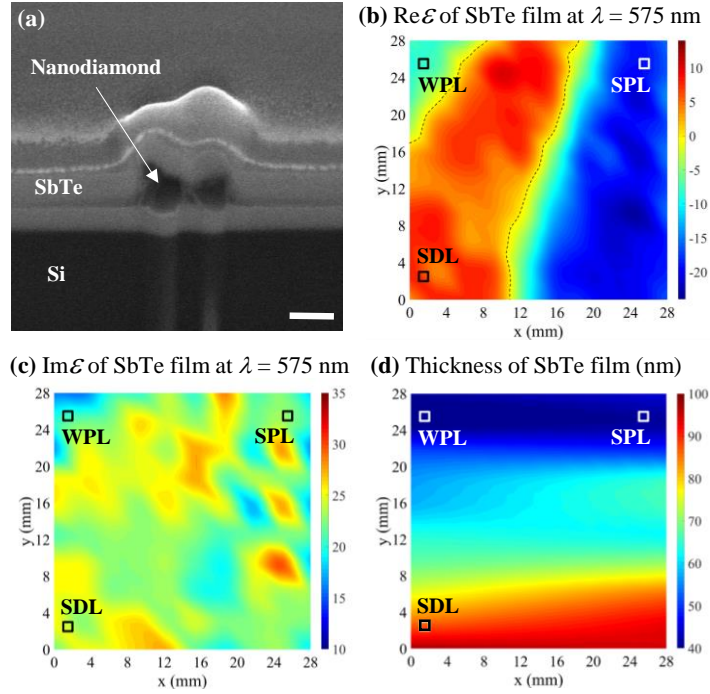

Figure 1. (a) SEM image of nanodiamonds embedded in SbTe films. Scale bar is 200 nm. (b) & (c) Real and imaginary parts of the permittivity of SbTe films mapped at the wavelength of 575 nm. Dashed curves in (b) trace Re $\varepsilon = 0$. (d) Inteprolated spatial map of the thickness of SbTe film, $d$, after two rounds of deposition. Squares in (b-d) mark the areas of the samples where the lifetime statistics of nitrogen-vacancy centers were collected: WPL - weakly plasmonic, SPL - strongly plasmonic, SDL - strongly dielectric.

The lifetimes of $NV^0$ centers were characterized using time-resolved cathodoluminescence (TR-CL). The measurements were performed at room temperature with a scanning electron microscope (SEM) operating at 10 kV in fixed-spot mode. An SEM beam blanker was driven with a wave function generator, which produced pulsed modulation of an electron beam and ensured the synchronization required for implementing time-correlated single-photon counting. In each TR-CL measurement, the beam current was maintained in the range 1.7 – 1.9 nA. The light emitted by the sample via cathodoluminescence was collected and collimated by a parabolic mirror. The collected light was then directed to the entrance slit of a VIS/NIR spectrometer Horiba iHR320, which allowed selecting photons in a 3.6 nm wide spectral window centerd at the zero phonon line. The filtered photons were registered with a single photon detector and the photon counts were arranged into time-correlated histograms (see figure 2(a)). The lifetimes of $NV^0$ centers were extracted by fitting the histograms with a bi-exponential function

$$I(t) = a_1\big(1 - \delta(t + a_2)\big) + 0.5 a_1 \delta(t + a_2)\big(e^{-\frac{t+a_2}{\tau_1}} + e^{-\frac{t+a_2}{\tau_2}}\big) + a_3, \quad (1)$$

where the fast exponential term corresponds to the carrier lifetime,[25] while the slow exponential term represents the decay rate of $NV^0$ centers.

The lifetime of $NV^0$ centers was characterized in three distinct areas of SbTe films, which are marked in figures 1(b)-1(d) by squares. They all had similar imaginary parts of the permittivity (at the level of ~20) and exhibited strong plasmonic response (SPL) with Re $\varepsilon = -15$, weak plasmonic response (WPL) with Re $\varepsilon = -7$ and strong dielectric response (SDL) with Re $\varepsilon = 7$, respectively. For each area, we selected a total of 15 ND clusters. The results of our measurements are summarized in figure 2(b) in the form of boxplots. Evidently, the reference sample featuring NDs on top of SbTe film displays a large variation of the lifetimes for all three areas with the spread exceeding 7 ns and the average lifetime of $NV^0$ centers nearing 20 ns, which is consistent with values reported in the literature.[21] At the same time, the average lifetimes of $NV^0$ centers in NDs incorporated by SbTe film are seen to become a factor of two shorter, converging at 11 ns for all three areas. More intriguingly, the spread of the lifetimes for this sample is reduced dramatically and becomes ~1 ns in the SPL and WPL areas, and ~2 ns in the SDL area. Similar variations are observed in figure 2(c) for the distributions of brightness of $NV^0$ centers, which we also characterised during TR-CL measurements. More specifically, the brightness data exhibit a wide spread for NDs placed on top of an SbTe film regardless of the local dielectric response of the film, but the spread becomes narrower by at least a factor of three for NDs incorporated into the SbTe film. The average brightness of $NV^0$ centers appears to also decrease, by about five times, for NDs embedded in the SbTe film, following the decrease of their average lifetime as revealed by figure 2(b).

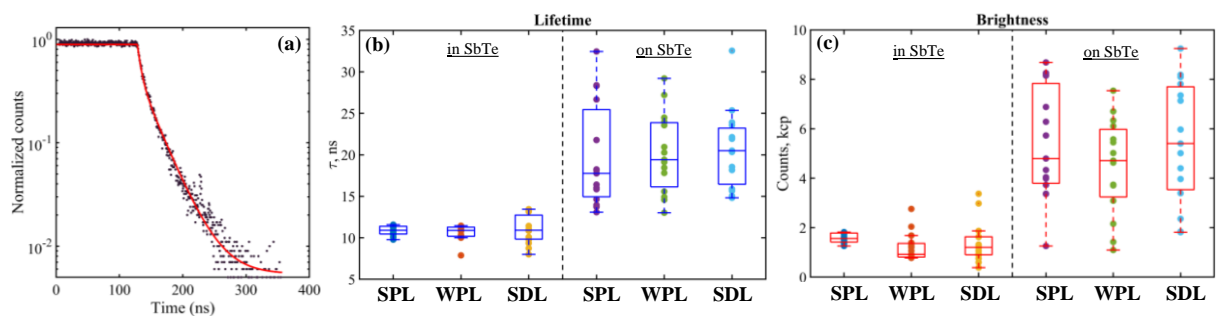

Figure 2. (a) An example of time-correlated histogram of photon counts due to cathodoluminescence of nitrogen-vacancy centers in a nanodiamond at the zero phonon line $\lambda = 575$ nm. Black dots correspond to measured data, red line represents a bi-exponential fit. (b) Boxplot of the distribution of lifetimes extracted from time-correlated histograms in areas of SbTe films with SPL (strong plasmonic), WPL (weak plasmonic) and SDL (strong dielectric) response, as marked in figures 1(b)-1(d). (c) Brightness of emission from nanodiamonds corresponding to the lifetime data in (b); kcp - kilocounts per second.

To understand the mechanism responsible for the observed changes in the lifetime statistics we modelled the emission of an isolated 120 nm large ND containing a single NV center with its electric dipole moment oriented parallel and orthogonal to the plane of the chalcogenide film. We considered two cases, as studied in our experiments, namely (i) an ND sitting on an SbTe film and (ii) an ND embedded in an SbTe film. Each case included films with WPL, SPL and SDL response, which were placed on a semi-infinite Si substrate. The modelling was performed in Comsol Multiphysics (see Supporting Information for details). Figure 3 presents the simulation results obtained for a lossy SbTe film featuring SDL response in terms of the energy absorption rate in various parts of the structure and electric field emitted into vacuum by an NV center with its electric dipole moment oriented parallel to the substrate. Evidently, for an ND embedded in an SbTe film, the overall energy absorption rate is higher and the areas featuring high absorption are seen to extend further along the film and into the Si substrate (compare figures 3(a) and 3(b)). Consequently, higher absorption should lead to stronger reduction of emission and that was exactly what we observed in our simulations (compare figures 3(c) and 3(d)). Similar differences in absorption and emission were observed for NDs on and in SbTe films featuring SPL and WPL response.

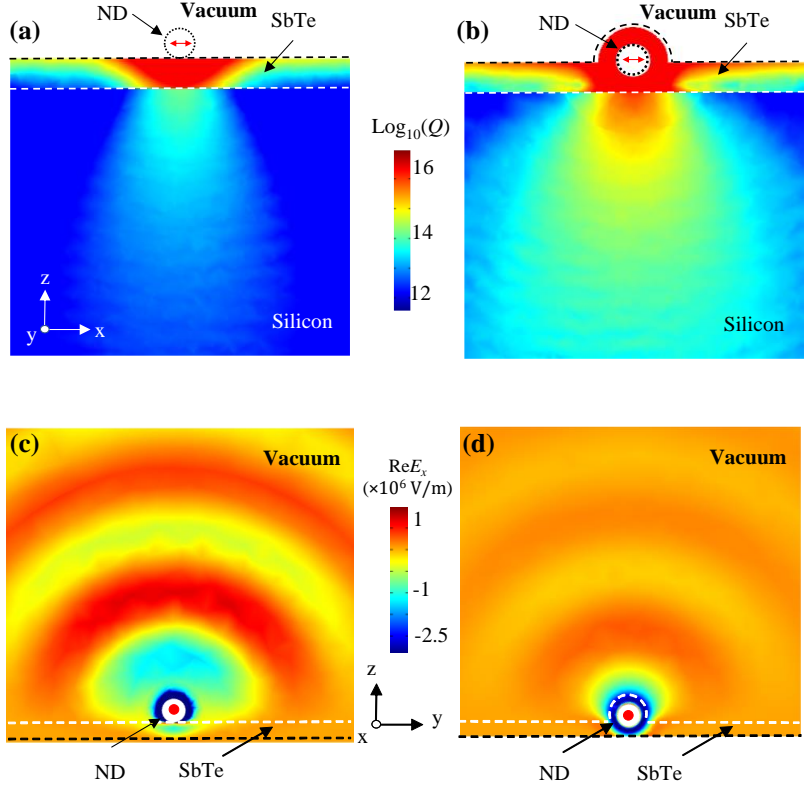

Figure 3. (a) & (b) Simulated distribution of absorption in the *xz*-plane presented in terms of rate of heat, *Q*, dissipated in the Si substrate and the SbTe film with a nanodiamond placed, respectively, on and in the film. The SbTe film here was modelled as a strong lossy dielectric (SDL) with $\varepsilon = 7 + i\,20$. The electric dipole moment of a nitrogen-vacancy center inside the nanodiamond was oriented parallel to the *x*-axis, as shown by the red arrows. (c) & (d) Simulated distribution of the *x*-component of the electric field (real part) in the *yz*-plane emitted into vacuum by the nitrogen-vacancy center inside the nanodiamond placed, respectively, on and in SbTe film. The electric field is not plotted inside the nanodiamond.

The simulated data enabled us to calculate the radiative and non-radiative decay rates for all possible configurations (see Supporting Information for details). The results are summarised and compared in Table S1. For the electric dipole moment of an NV center oriented perpendicular to the substrate, the non-radiative decay rate, $\gamma_{\text{non}}^{\perp}$, is about 4.5 times higher when an ND is embedded in as opposed to deposited on the SbTe film (for all three types of SbTe, SPL, WPL, and WDL). Expectedly, the radiative decay rate, $\gamma_{\text{rad}}^{\perp}$, in this case is lower – by a factor of 2 for plasmonic SbTe films and by a factor of 9 for a strongly dielectric SbTe film. When the electric dipole moment is parallel to the substrate the situation is similar, though the differences in the decay rates, $\gamma_{\text{non}}^{\parallel}$ and $\gamma_{\text{rad}}^{\parallel}$, appear to be larger: $\gamma_{\text{non}}^{\parallel}$ is > 18 times higher and $\gamma_{\text{rad}}^{\parallel}$ is > 3 times lower for all three types of an SbTe film incorporating an ND. We further note that the ratio between the non-radiative and radiative decay rates does not exceed 2 for an ND sitting on an SbTe film, but raises above 9 for an ND embedded in an SbTe film (approaching 174 when SbTe behaves as a strong dielectric and the electric dipole moment is parallel to the substrate). Our analysis, therefore, indicates that the experimentally observed narrowing of the lifetime distribution was enforced by substantially increased non-radiative decay, which completely dominates the relaxation of NV centers in embedded NDs. The enhancement of non-radiative decay should lead to shortening of the average lifetime in accordance with our experimental observations (see figure 2(b)).

The calculated data in table S1 also suggests that there is another mechanism responsible for changes in the lifetime statistics – an interplay between scattering and absorption of emitted photons. For an ND embedded in an SbTe film, it renders the total decay rate, $\gamma_{tot}$, of an NV center almost insensitive to the orientation of its electric dipole moment ($\gamma_{tot}^{\perp}/\gamma_{tot}^{\parallel} \sim 1$). Thus, if such an ND contains many NV centers (as our samples) this mechanism will efficiently negate the variation of lifetimes across an ND due to random orientations of the electric dipole moments.[21] On the contrary, for an ND sitting on an SbTe film the dependence of $\gamma_{tot}$ on the electric dipole alignment remains quite strong ($\gamma_{tot}^{\perp}/\gamma_{tot}^{\parallel} > 3.5$) and, therefore, can only lead to further spreading of the lifetime distribution (especially when SbTe exhibits plasmonic response).

In conclusion, we have experimentally demonstrated that embedding nanodiamonds with multiple nitrogen-vacancy centers in thin chalcogenide films enabled us to dramatically narrow the distribution of the decay rates and shorten the average lifetime of the fluorescent defects. Our numerical analysis has shown that the observed changes in the lifetime statistics, i.e., more than five-fold reduction of the spread and two-fold decrease of the average, were underpinned by two mechanisms. One of the mechanisms is dissipation in lossy chalcogenide films, which enhances non-radiative decay to the extent that it completely dominates radiative decay for nanodiamonds embedded in the films. While such a mechanism naturally reduces the brightness of a single nitrogen-vacancy center, in practice the intensity of fluorescence (and, in our case, cathodoluminescence) can be maintained at a sufficient level in nanodiamonds containing many centers. The other mechanism corresponds to an interplay between scattering and absorption of emitted photons. For embedded nanodiamonds it renders the total decay rate of nitrogen-vacancy centers insensitive to the orientation of their electric dipole moments, which is random in large ensembles of the defects. Given that one can dial (via stoichiometric engineering) optical response of chalcogenides anywhere from strongly plasmonic to strongly dielectric,[22] and also can vary the level of losses by switching (optically, electrically or thermally) between amorphous and crystalline phases,[26-28] our approach offers a straightforward way of implementing active control over the emission statistics in nanodiamonds.

## Acknowledgements

The authors would like to acknowledge the financial support of the UK's Engineering and Physical Sciences Research Council (Grants No. EP/M0091221), China Scholarship Council (No. 201608440362). Following a period of embargo, the data from this paper will be available from the University of Southampton ePrints research repository: http://doi.org/10.5258/SOTON/DXXX.

# Supporting Information

# Engineering the Collapse of Lifetime Distribution of Nitrogen-Vacancy Centers in Nanodiamonds

H. Li[†,*], J. Y. Ou[†], B. Gholipour[†], J. K. So[†,‡], D. Piccinotti[†], V. A. Fedotov[†], N. Papasimakis[†]

[†] *Optoelectronics Research Center & Center for Photonic Metamaterials, University of Southampton, Southampton SO17 1BJ, UK*

[‡] *Center for Disruptive Photonic Technologies, School of Physical and Mathematical Sciences and The Photonics Institute, Nanyang Technological University, Singapore 637371, Singapore*

**Calculating Lifetimes of an Nitrogen-Vacancy Center in a Nanodiamond**

Numerical simulations were performed using the RF module of COMSOL 5.3a, a commercial simulation software based on the finite-element method. The simulation domain of our model is shown in figure S1. It is represented by a cylinder with a diameter of 1.5 μm and a height of 3 μm. All surfaces of the cylinder were terminated with scattering boundaries. A nanodiamond (ND) was modelled as a 120 nm large dielectric sphere with refractive index 2.40.[29] The refractive index of the Si substrate was set to $4.00 + i\,0.03$.[30] A nitrogen-vacancy (NV) center radiating at the wavelength of 575 nm was introduced as a volume polarization density oscillating inside a sphere with radius $r_s = 5$ nm, which was placed in the center of the ND. The direction of the polarization (and, therefore, the orientation of the effective electric dipole moment) was set either parallel or perpendicular to the substrate.

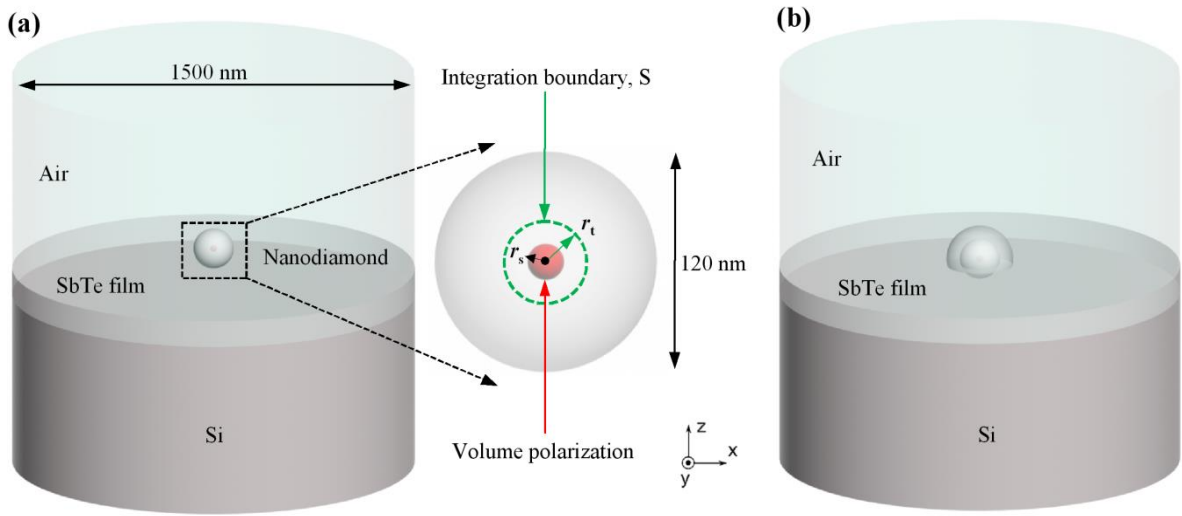

**Figure S1.** A schematic of the simulation domain used for modelling the emission of a nanodiamond containing a single NV center with the nanodiamond placed either on an SbTe film (a) or embedded in an SbTe film (b).

The total power emitted by the NV center, $P_{rad}$, was calculated by integrating the power flow over the surface of a sphere with a radius $r_t = 6$ nm, which encapsulated the emitting NV center. The lifetime of the NV center was obtained as $\tau = \frac{\tau_{bulk}P_{bulk}}{P_{rad}}$, where $P_{bulk}$ is the total power emitted by the NV center inside bulk diamond and $\tau_{bulk}$ corresponds to the known bulk lifetime $\tau_{bulk} = 19$ ns.[31] $P_{bulk}$ was calculated in the same way as $P_{rad}$ except that the entire simulation domain was set as diamond. The total decay rate was calculated as $\gamma_{tot} = \frac{1}{\tau} = \frac{P_{rad}}{\tau_{bulk}P_{bulk}}$. The radiative decay rate was given by $\gamma_{rad} = \frac{P_0}{\tau_{bulk}P_{bulk}}$, where $P_0$ is the power radiated into the far-field zone above the substrate (since the substrate was lossy, the power radiated into the substrate was eventually absorbed). The non-radiative decay rate was calculated as $\gamma_{non} = \gamma_{tot} - \gamma_{rad}$. Table S1 compares the decay rates of the NV center calculated for the two orientations of its electric dipole moment and different types of SbTe optical response.

**Table S1.** Calculated decay rates of a single NV center interacting with thin SbTe films, which exhibit weak plasmonic (WPL), strong plasmonic (SPL) and strong dielectric (SDL) response.

| Decay rates | WPL (Re $\varepsilon = -7, d = 40$nm) | | SPL (Re $\varepsilon = -15, d = 40$nm) | | SDL (Re $\varepsilon = 7, d = 95$nm) | |
|---|---|---|---|---|---|---|
| | on SbTe | in SbTe | on SbTe | in SbTe | on SbTe | in SbTe |
| $\gamma_{tot}^{\parallel}$ (µs$^{-1}$) | 8.97 | 102.79 | 8.38 | 114.36 | 8.64 | 88.96 |
| $\gamma_{tot}^{\perp}$ (µs$^{-1}$) | 36.21 | 99.58 | 37.64 | 103.99 | 31.47 | 90.30 |
| $\gamma_{non}^{\parallel}$ (µs$^{-1}$) | 4.05 | 101.16 | 3.12 | 112.87 | 4.86 | 88.45 |
| $\gamma_{non}^{\perp}$ (µs$^{-1}$) | 20.13 | 90.97 | 17.25 | 93.74 | 20.07 | 89.01 |
| $\gamma_{rad}^{\parallel}$ (µs$^{-1}$) | 4.91 | 1.62 | 5.26 | 1.49 | 3.78 | 0.51 |
| $\gamma_{rad}^{\perp}$ (µs$^{-1}$) | 16.08 | 8.61 | 20.39 | 10.25 | 11.40 | 1.28 |
| $\gamma_{non}^{\parallel}/\gamma_{rad}^{\parallel}$ | 0.83 | 62.29 | 0.59 | 75.78 | 1.29 | 173.63 |
| $\gamma_{non}^{\perp}/\gamma_{rad}^{\perp}$ | 1.25 | 10.57 | 0.85 | 9.14 | 1.76 | 69.42 |